\documentclass[draft]{agujournal2019}
\usepackage{url} 
\usepackage{lineno}
\usepackage[inline]{trackchanges} 
\usepackage{soul}
\usepackage{amsmath}

\justifying
\usepackage{float}

\draftfalse

\journalname{JGR: Machine Learning and Computation}

\begin{document}

\title{Machine Learning-based Denoising of Surface Solar Irradiance simulated with Monte Carlo Ray Tracing
}

\authors{M. Reeze\affil{1}, M. A. Veerman\affil{1}, C. C. van Heerwaarden\affil{1}}

\affiliation{1}{Meteorology and Air Quality Group, Wageningen University \& Research, Wageningen, The Netherlands}

\correspondingauthor{M. A. Veerman}{menno.veerman@wur.nl}

\begin{keypoints}
\item We train a denoising autoencoder for Monte Carlo ray tracing estimates of surface solar irradiance fields below cumulus clouds

\item Denoising diffuse irradiance exceeds accuracy of higher sampling budgets that require over tenfold the computational cost 

\item Denoising direct irradiance is effective in sunlit areas but significant noise remains near cloud shadow edges

\end{keypoints}

\begin{abstract}
Simulating radiative transfer in the atmosphere with Monte Carlo ray tracing provides realistic surface irradiance in cloud-resolving models. However, Monte Carlo methods are computationally expensive because large sampling budgets are required to obtain sufficient convergence. Here, we explore the use of machine learning for denoising direct and diffuse surface solar irradiance fields. We use Monte Carlo ray tracing to compute pairs of noisy and well-converged surface irradiance fields for an ensemble of cumulus cloud fields and solar angles, and train a denoising autoencoder to predict the well-converged irradiance fields from the noisy input. We demonstrate that denoising diffuse irradiance from 1 sample per pixel (per spectral quadrature point) is an order of magnitude faster and twice as accurate as ray tracing with 128 samples per pixel, illustrating the advantage of denoising over larger sampling budgets. Denoising of direct irradiance is effective in sunlit areas, while errors persist on the edges of cloud shadows. For diffuse irradiance, providing additional atmospheric information such as liquid water paths and solar angles to train the denoising algorithm reduces errors by approximately a factor of two. Our results open up possibilities for coupled Monte Carlo ray tracing with computational costs approaching those of two-stream-based radiative transfer solvers, although future work is needed to improve generalization across resolutions and cloud types.
\end{abstract}

\section*{Plain Language Summary}
As atmospheric models move towards higher resolutions and the demand for solar energy forecasts increases, there is an urgent need for accurate modeling of solar radiation. One of the most accurate techniques to simulate the transfer of radiation through the atmosphere is Monte Carlo ray tracing. However, this technique requires many computational resources, limiting its application in operational contexts. This problem has been extensively researched in the movie and gaming industry, where state-of-the-art solutions use machine learning algorithms to approximate illumination in artificial scenes. Here, we explore the use of these algorithms in the context of atmospheric modeling, with a focus on incoming solar radiation under broken clouds. Using this approach, we are able to deliver close approximations of realistic surface irradiance at a fraction of the original computational cost.

\section{Introduction}
Monte Carlo ray tracing is one of the the most accurate techniques for simulating atmospheric radiative transfer \cite{Cahalan2005}. Monte Carlo ray tracing can deliver surface irradiance patterns closely resembling observations, capturing the characteristic bimodal distribution under cumulus clouds and associated areas with cloud enhancements \cite{Gristey2020a,He2024}. Recently, \citeA{Veerman2022} and \citeA{Tijhuis2024} performed large-eddy simulations (LES) of shallow cumulus clouds with a coupled Monte-Carlo ray tracer, showing that these irradiance patterns can have strong consequences for cloud evolution, with locally enhanced surface heat fluxes resulting in thicker and wider clouds. 

Monte Carlo ray tracing requires large sampling budgets (number of samples per pixel) to obtain sufficiently converged irradiance fields, which is computationally expensive and often limits its application in operational contexts \cite{He2024, Gristey2020b, Villefranque2023}. While performance can be improved by constructing acceleration grids \cite{Villefranque2019}, or by utilizing the computing capabilities of modern graphics processing units (GPUs) \cite{Veerman2022}, Monte Carlo radiative transfer methods remain significantly more expensive than widely used two-stream approximations. The search for computationally efficient Monte Carlo ray tracing has been more extensive in the movie and video game industry \cite<e.g.,>[]{Chaitanya2017, Dahlberg2019, Huo2021, Isik2021}. There, solutions to reduce computational costs of Monte Carlo ray tracing are divided into two main categories: advanced sampling strategies and post-processing reconstruction schemes \cite{Huo2021}. The focus of our study will be on the latter, also known as denoising. 

Classical denoising methods comprise various filtering techniques \cite{Fan2019}. Examples include Gaussian, bilateral \cite{Tomasi1998}, non-local means \cite{buades2005}, total variation \cite{Rudin1992}, and wavelet \cite{Chang2000} filters. Although such classical denoising methods perform reasonably well and are relatively simple to implement, they come with drawbacks such as edge blurring and loss of high frequency information \cite{Fan2019, Tian2020}. In recent years, the application of machine learning (ML) algorithms for removing Monte Carlo noise has led to notable advancements \cite{Huo2021, Tian2020}. The principle of ML-based denoising is to use a noisy image and a set of auxiliary variables as inputs for an algorithm which learns to produce an approximation of the converged image. State-of-the-art ML-based denoising methods are typically based on Convolutional Neural Networks (CNNs) \cite{Bako2017,Fan2019,Huo2021,Tian2020,Zhang2017}. In particular denoising autoencoders, advanced variants of CNNs, have proven effective in denoising of Monte Carlo scenes with low sampling budgets and high corresponding noise levels \cite{Chaitanya2017,Kuznetsov2018,Mao2016}.

The application of machine learning to 3D radiative transfer problems is not new in the atmospheric sciences. 
For example, \citeA{Masuda2019} trained CNNs to predict cloud optical depths from sky-view images generated synthetically with 3D Monte Carlo radiative transfer. 
\citeA{Yang2022} used CNNs to estimate aerosol optical depths from reflectance fields derived from 3D radiative transfer computations. 
\citeA{Gristey2020b} and \citeA{gristey2022} used machine learning to predict probability density functions of surface solar irradiance fields, produced with Monte Carlo radiative transfer, from cloud field and aerosol properties. 
\citeA{Meyer2022} trained neural networks to predict subgrid 3D cloud radiative effects from the SPARTACUS \cite<SPeedy Algorithm for Radiative TrAnsfer through CloUd Sides;>{Hogan2016} solver.

In this study, we explore the potential of denoising autoencoders for speeding-up surface irradiance computations based on Monte Carlo ray tracing. 
By doing so, we aim to improve the affordability of Monte Carlo radiative transfer in atmospheric models.
We investigate the optimal sampling budget of the initial Monte Carlo estimate, which is a trade-off between speed and accuracy. A larger sampling budget provides more information to the denoising algorithm, but increases computational cost. Furthermore, we investigate the added value of providing auxiliary input variables to the autoencoder, such as liquid water paths and solar angles. Consequently, the main question of this paper is:

\textit{How well can an autoencoder denoise ray-traced surface irradiance?}

We define the following subquestions to provide a quantitative answer to the main question:
\begin{itemize}
    \item How does the sampling budget affect the trade-off between speed and accuracy?
    \item How do auxiliary inputs contribute to prediction accuracy?
\end{itemize}

Surface solar irradiance is generally considered as the sum of a direct and diffuse component, where the diffuse component is all radiation scattered at least once within the atmosphere. In this study, we choose to denoise direct and diffuse irradiance independently, since these components present distinct noise patterns. Additionally, we focus on the irradiance beneath shallow cumulus clouds, which are common over large parts of the world \cite{Berg2011}, are strongly coupled to the earth's surface and produce complex spatial irradiance patterns \cite{Burleyson2015,Gristey2020a,He2024}. 

In section 2, we introduce the methodology regarding generation of training data, network architecture and model training. Section 3 presents the results, discussing the sampling budget trade-off and impact of auxiliary variables. In section 4, we summarize the main conclusions and discuss their implications.

\section{Methods}

\subsection{Generation of Training Data}

\begin{figure}[H]
\noindent\includegraphics[width=\textwidth]{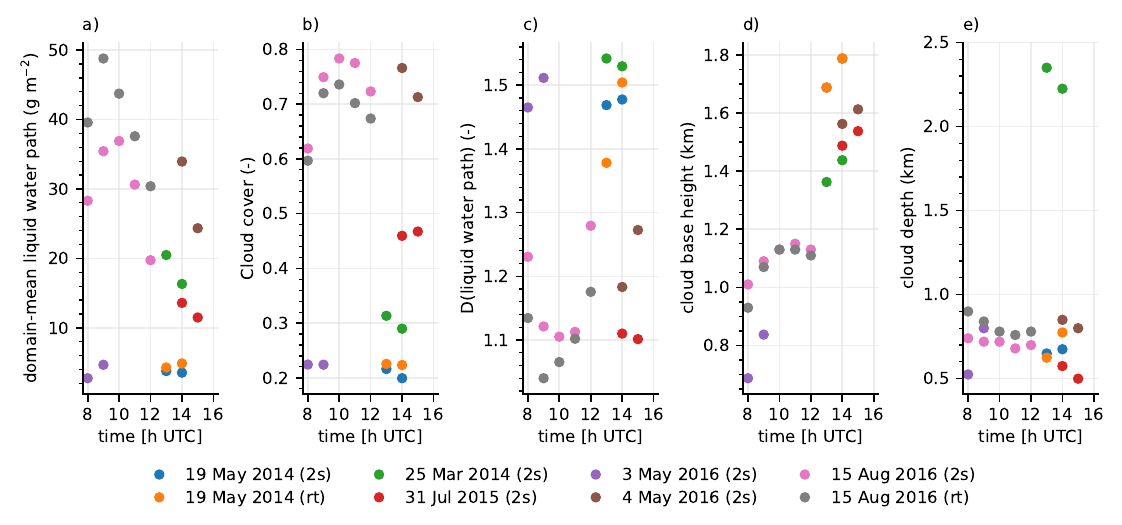}
\caption{(a) domain-mean liquid water path, (b) cloud cover, (c) relative dispersion of liquid water path (standard deviation $/$ mean) following \citeA{Gristey2020a}, (d) cloud base height and (e) cloud depth, for all cloud fields used to generate training data. Colors denote the different simulations the cloud fields are taken from. Cloud depth is the height difference between the lowest and highest layers where liquid water path $>$ 0 g m$^{-2}$.}
\label{cloud_field_summary}
\end{figure}

We generated a dataset using 22 shallow cumulus cloud fields, 10 from the large-eddy simulations of \citeA[5 each from their \texttt{rt} and \texttt{2s} experiments]{Veerman2022} and 12 from additional large-eddy simulations based on a subset of the cumulus days selected by \citeA{Tijhuis2024}: 4 each from \texttt{25 Mar 2014}, \texttt{31 Jul 2015}, \texttt{03 May 2016} and \texttt{04 May 2016} (large-eddy simulations performed with 1D radiative transfer) and 2 each from two different \texttt{19 Mar 2014} large-eddy simulations (one with 1D, one with 3D radiative transfer). These days were selected such that they span a wide range of cloud covers and cloud base heights (Figure \ref{cloud_field_summary}). All simulations are performed with the MicroHH large-eddy simulation code \cite{vanheerwaarden2017}, with a horizontally periodic domain of 38.4 $\times$ 38.4 km$^{2}$ and a horizontal grid spacing of 50 m, although the cloud fields are subsequently coarsened to 100 m resolution by horizontally averaging adjacent grid cells, which reduces the computational costs of the ray tracing computations. The vertical extent of the domain is 4 km with a grid spacing of 20 m in the simulations from \citeA{Veerman2022} and 6.4 km with a grid spacing of 25 m in the simulations based on \citeA{Tijhuis2024}. For each cloud field, we sampled 10 combinations of solar zenith and azimuth angles. We used a maximum zenith angle of 80°: ray tracing costs increase drastically for larger zenith angles and shallow cumulus rarely co-occur with larger zenith angles \cite<e.g.,>[]{Gristey2020a}. Broadband solar radiative fluxes are computed with the forward Monte Carlo radiative transfer solver presented by \citeA{Veerman2022} and further described by \citeA[chapter 3]{veerman2023thesis}.
We use the set of 112 shortwave quadrature points of RTE+RRTMGP \cite<Radiative Transfer for Energetics + RRTM for General circulation models$-$Parallel;>[]{Pincus2019} for spectral integration.

The cloud fields, irradiance fields and azimuth angle in the dataset are subsequently rotated 90°, 180° and 270°. This form of data augmentation increases diversity of the dataset, helping to prevent overfitting. The sampling of angles and the rotation results in a final dataset consisting of 880 distinct direct and diffuse irradiance fields, which we found to be sufficiently large for the aim of this study.

We used irradiance fields with 4096 samples (photons) per pixel per spectral quadrature point (spp) as targets for training our model.  For each pixel, this amounts to a total of 458,752 samples, equally distributed over the 112 spectral quadrature points. With 4096 spp, the remaining uncertainty of the Monte Carlo estimate for the surface solar irradiance is only about 2 W m$^{-2}$ ($<$ 0.5\% of the mean surface irradiance).
The order of magnitude of this uncertainty is smaller than the error of solar irradiance measurement devices \cite<e.g.,>{sensor2004,Mol2024}. Additionally, \citeA{Veerman2022} showed that beyond 32 spp, convergence of the ray tracer hardly affects the simulated behavior of the atmosphere. For these reasons, we argue that using 4096 spp for target irradiance suffices for the aim of this study.

The dataset is split up into a training (70\%), validation (15\%) and testing (15\%) dataset. The training dataset is used to train the models, the validation dataset is used to tune hyperparameters and analyze overfitting, and the testing dataset is used to evaluate performance of the final models. 

\subsection{Network Architecture}
\begin{figure}[H]
\noindent\includegraphics[width=\textwidth]{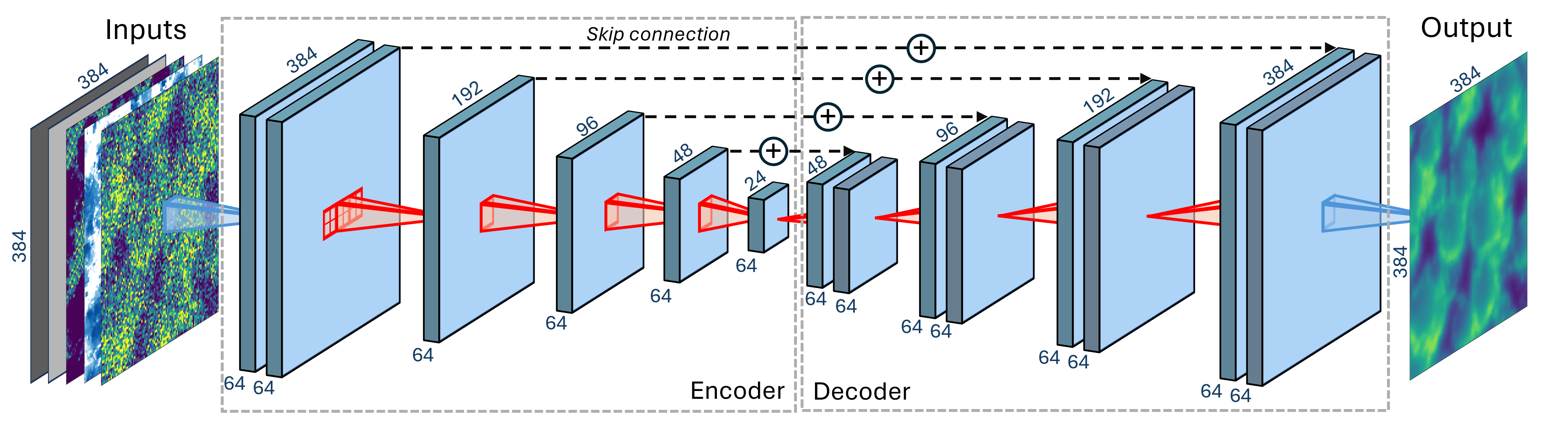}
\caption{Schematic overview of our autoencoder architecture. The input consists of the solar zenith and azimuth angle (both uniform), noisy direct and diffuse irradiance, and the liquid water path. The output in this example is an approximation of the converged diffuse irradiance field. The red filters visualize max pooling (maximum value within the pooling filter) in the encoder and nearest-neighbor upsampling in the decoder, both using a filter size of 2 $\times$ 2 and a stride of 2 as visualized at the first hidden layer (red squares). The blue filters visualize convolutions in the outer layers. The hidden layers also perform convolutions, resulting in the features maps with dimension as shown for each hidden layer. Note that the shown filters are not to scale, and example input and output fields are zoomed-in. Spatial dimensions of the input, output and hidden layers during training are shown, while the trained model can be applied to domains with different dimensions. Design is based on \citeA{Chaitanya2017}.}
\label{AEplot}
\end{figure}

Although our goal is denoising irradiance fields instead of RGB images, we approach it as a Monte Carlo-rendered image denoising problem, for which CNNs and especially convolutional denoising autoencoders have already been proven successful \cite<e.g.>{Fan2019, Chaitanya2017}.
We use an autoencoder architecture similar to the one presented by \citeA{Chaitanya2017}, see Figure \ref{AEplot} for an overview of the architecture. 
The encoder part of the network uses a series of convolutional and max pooling layers (for feature extraction and down-sampling, respectively) to compress the noisy input data into a latent space representation.
The decoder then uses a series of convolutional and upsampling (nearest-neighbor) layers to reconstruct denoised irradiance fields from the latent space representation.  
Information is also passed between the encoder and decoder through skip connections (Figure \ref{AEplot}), allowing for deeper learning while preventing a loss in detail \cite{He2016, Mao2016}. 
Since we do not denoise in a temporal context, we omit the recurrent connections of the network of \citeA{Chaitanya2017}.

Our denoising autoencoder uses 5 input fields to predict 1 output field (Figure \ref{AEplot}). The input fields include the noisy direct and diffuse irradiance with a low sampling budget (1 - 32 spp). The noisy direct irradiance can be viewed as an auxiliary variable for the model with converged diffuse irradiance as target and vice versa. Both the input and target direct and diffuse irradiance are normalized with the cosine of the solar zenith angle, which slightly improves accuracy ($\approx$ 1 W m$^{-2}$ lower RMSE). The other auxiliary variables consist of the solar zenith angle, solar azimuth angle, and the natural logarithm of the vertically summed liquid water path (LWP). To avoid the undefined logarithm of zero, we add a constant ($\epsilon$ = 1) to the LWP before computing the logarithm. The inputs and targets are subsequently normalized with their mean and standard deviation over the full dataset. 

The first encoder stage uses two convolutional layers, while the other encoder stages only apply one convolution (Figure \ref{AEplot}). In the decoder, the skip connections add the feature maps of the encoder element-wise to the upsampled feature maps using leaky rectified linear unit (Leaky ReLU) activation \cite{Maas2013} with $\alpha$ = 0.05 to introduce non-linearity. After the skip connections, each decoder stage applies two sets of convolution. The last decoder stage uses an extra convolutional layer to obtain the final output. 
All convolutional layers use a kernel size of 3 and a stride of 1, also with leaky ReLU activation ($\alpha$ = 0.05). To preserve spatial dimensions in the convolutional layers we use circular padding, which is consistent with our periodic cloud and irradiance fields. Each hidden convolutional layer generates the same number of feature maps (Figure \ref{AEplot}). While \citeA{Chaitanya2017} increase the number of channels for layers with smaller spatial dimensions, we found that this did not significantly improve accuracy for our denoising problem. 

Our autoencoder uses residual learning, which means that the network is trained to predict the noise of the original input \cite<e.g.,>{Zhang2017, Mao2016}. This noise is consequently subtracted from the original input to obtain a clean output. To avoid the model predicting negative irradiance values, we add a manual activation function after subtracting the residuals from the original input. This function disallows values below the normalized target value for zero irradiance: $ \max(-\frac{\overline{y}}{\sigma_y}, x) $, where $\overline{y}$ and $\sigma_y$ are the mean and standard deviation that have been used to normalize the target irradiance, and $x$ is the model prediction of the last convolutional layer. 

For an optimum balance between accuracy and training feasibility, we use 4 pooling layers and 64 channels in the hidden layers (Figure \ref{AEplot}). Further increasing the number of pooling layers or channels improves performance only marginally, at the expense of increasing training costs. This autoencoder architecture has a receptive field of 128 $\times$ 128 pixels, meaning that each output pixel processes and responds to this specific region of the input. The architecture results in 483,585 trainable parameters with a total size of 1.9 MB. The autoencoder allocates 290 MB peak memory to denoise a field with 384 $\times$ 384 pixels.

\subsection{Network Training}
Our networks are trained for 250 epochs, using the Adam optimizer \cite{Kingma2017} and an initial learning rate of $7\cdot10^{-4}$, multiplied with a factor of 0.7 after each 25th epoch. We train the models using the entire domains of 384 $\times$ 384 pixels in the training dataset. The fully convolutional nature of the network allows the trained network to be applied to inputs with different dimensions, although the way we set up the skip connections restricts the input spatial dimensions to multiples of 16 (i.e. 4 $\times$ number of pooling layers). The order of the training data is randomized each epoch and processed in batches of 8.

As loss function, we used the Mean Squared Error (MSE) plus a bias term to encourage conservation of mean surface irradiance:
\begin{equation}
    \textit{Loss} = \textit{MSE} + \beta \cdot \textit{SME}.
\end{equation}
Here, SME is the Squared Mean Error and $\beta $ is a parameter that determines the importance of this term relative to the MSE. This approach is inspired by \citeA{Beucler2021}, who showed the effectiveness of penalizing a neural network for violating physical constraints with the loss function. We use $\beta = 5 $ to train our networks, which we found to improve mean irradiance conservation without compensating in RMSE. Separate networks have to be trained for each sampling budget, because noise levels are different. For example, applying the autoencoder trained on 4 spp to noise of 1 or 16 spp leads to more than a doubling of the RMSE compared to autoencoders trained specifically on 1 or 16 spp.

\subsection{Evaluation Methods}
In addition to the Root Mean Squared Error (RMSE) and Mean Absolute Error (MAE) we use 1$-$SSIM, where SSIM is the Structural Similarity Index Measure, as a metric of distortion. The SSIM is calculated based on local averages, local standard deviations and local covariance between predictions and targets \cite{wang2004}. To assess biases, we use the Mean Error (ME). 
Furthermore, we compute radially-averaged power spectral densities of the irradiance fields to assess the performance of the denoiser at different spatial scales.
We evaluate the runtimes of the autoencoder and ray tracer using a NVIDIA H100 Graphical Processing Unit (GPU). Discussed ray tracer runtimes do not include the time spent on computing optical properties and are for both direct and diffuse irradiance, as both irradiance components are estimated simultaneously.
Besides comparing the denoised predictions to the target of 4096 spp, we also compare the predictions to a reference of 128 spp. Albeit being hardly noise free, the level of convergence reached with 128 spp is currently deemed appropriate, in terms of accuracy and computational costs, for large-eddy simulations with coupled ray tracing \cite<e.g.,>{Veerman2022, Tijhuis2024}. We compute the performance metrics of both the denoising approach and the 128 spp reference with respect to the 4096 spp target.

\section{Results}

\subsection{Visualizing Denoising Performance}

\begin{figure}[h]
\noindent\includegraphics[width=\textwidth]{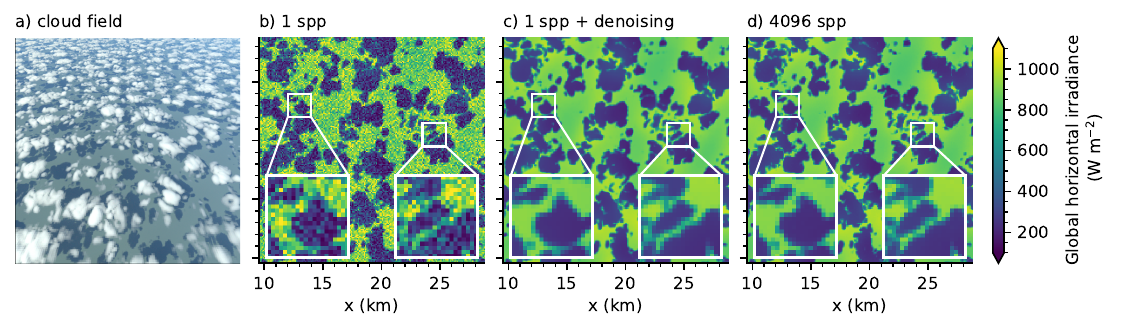}
\caption{(a) Visualization of a shallow cumulus cloud field (rendered with the backward ray tracer described by \citeA[Chapter 3]{veerman2023thesis}) and corresponding global horizontal irradiance (GHI): (b) input with 1 sample per pixel (spp); (c) denoised input using the autoencoder; (d) target with 4096 spp. GHI figures are zoomed-in on the middle quarter of the domain (x = 9.6 $-$ 28.8 km, y = 9.6 $-$ 28.8 km). Solar zenith angle = $ 45^\circ $, solar azimuth angle = $ 210^\circ $.}
\label{cloudfield_ghi_plot}
\end{figure}

\begin{figure}[H]
\noindent\includegraphics[width=\textwidth]{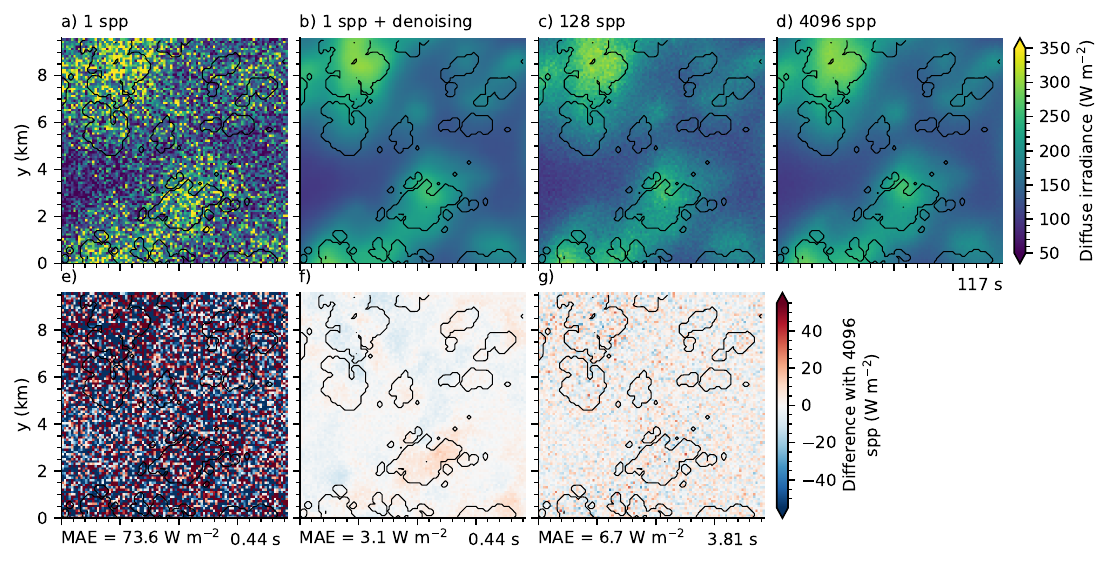}
\caption{Diffuse irradiance field (top row) and differences compared to target (bottom row): (a,e) input with 1 sample per pixel (spp); (b,f) 1 spp input denoised with the autoencoder; (c,g) reference with 128 spp; (d) target with 4096 spp. Figure is zoomed in on a 16\textsuperscript{th} of the domain (x = 0 $-$ 9.6 km, y = 0 $-$ 9.6 km) to highlight differences. Mean Absolute Error (MAE) and runtimes (bottom right) represent the entire domain. Contours represent cloud outlines, where liquid water path (LWP) $ > 0 \  \text{kg} \, \text{m}^{-2} $. Solar zenith angle = $ 36^\circ $, solar azimuth angle = $ 209^\circ $.}
\label{dif_fields}
\end{figure}

\begin{figure}[h]
\noindent\includegraphics[width=\textwidth]{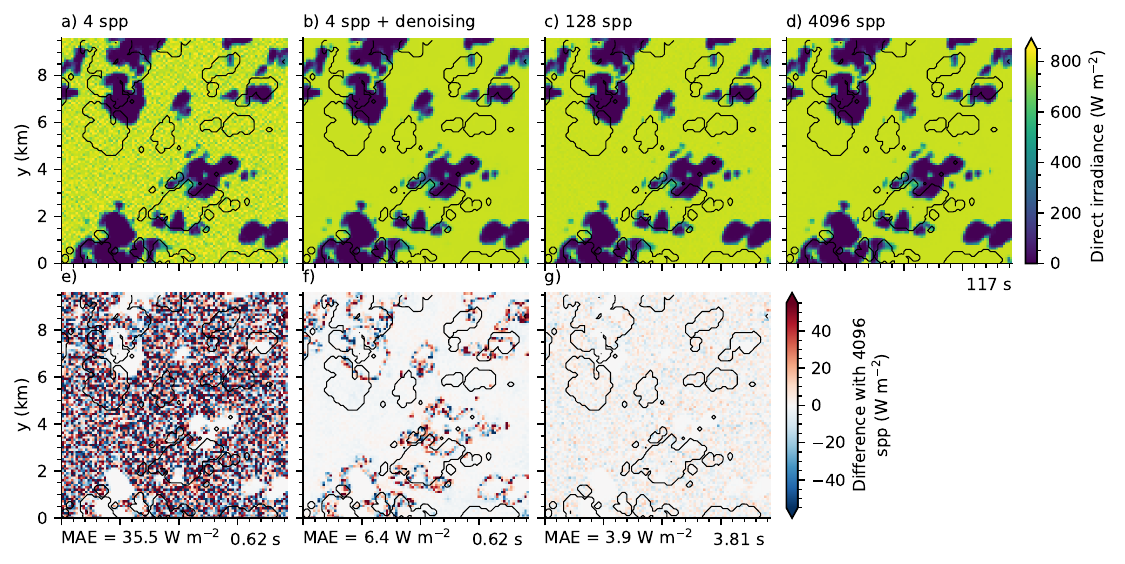}
\caption{As Figure \ref{dif_fields}, but for direct irradiance and 4 samples per pixel.}
\label{dir_fields}
\end{figure}

Figure \ref{cloudfield_ghi_plot} illustrates the performance of the denoising algorithms for an example shallow cumulus cloud field. When ray tracing with only 1 spp, large-scale patterns in the global horizontal irradiance (GHI), such as cloud shadows and regions with strong cloud enhancements, are already quite well resolved. However, the GHI field still shows severe noise on smaller scales, with local errors of O(100) $\mathrm{W\ m^{-2}}$. To the eye, the denoised and target irradiance are nearly indistinguishable. The GHI is obtained by denoising direct and diffuse irradiance separately, which is more extensively discussed below. We start by demonstrating denoising performance at 1 spp for diffuse irradiance and at 4 spp for direct irradiance (Figure \ref{dif_fields} and \ref{dir_fields}). We choose to visualize performance for these sampling budgets based on the analysis in Section 3.2.

The denoising method outperforms the 128 spp reference method for diffuse radiation (Figure \ref{dif_fields}). The diffuse irradiance with 128 spp has more noise than the denoised field and double the MAE. The noise in input (1 spp) and reference (128 spp) irradiance is larger in cloud shadows than in sunlit areas (Figure \ref{dif_fields}e,g), while the remaining errors in denoised diffuse irradiance do not seem to correlate with the larger irradiance patterns (Figure \ref{dif_fields}f). Although the remaining errors are small, a possible downside of the denoising approach is the larger spatial scale of the error patterns (Figure \ref{dif_fields}f). The 1 spp + denoising method is an order of magnitude faster than the 128 spp reference, and two orders of magnitude faster than the 4096 spp target (Figure \ref{dif_fields}). The runtime of the denoising autoencoder ($ \approx $ 2 ms) is negligible compared to the runtime of the radiative transfer scheme.

Noise in direct irradiance is predominantly present in sunlit areas, while cloud shadows are mostly free of noise (Figure \ref{dir_fields}e,g). The denoising approach results in less noise in the sunlit areas compared to the 128 spp reference, but larger errors remain on the edges of cloud shadows (Figure \ref{dir_fields}f,g). Apparently, the denoiser is unable to leverage the information on solar zenith and azimuth angles to project the vertically integrated liquid water path to exactly the right pixels on the surface. Nevertheless, errors at the shadow edges are relatively small ($ \approx $ 0-5\%) compared to the order of magnitude of direct irradiance.

\begin{figure}[!h]
\noindent\includegraphics[width=\textwidth]{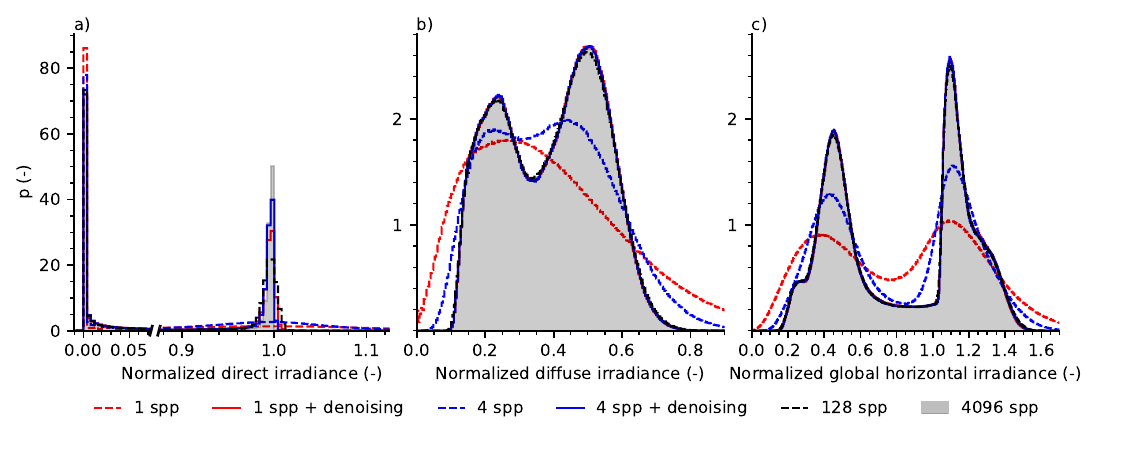}
\caption{Probability density functions of (a) direct, (b) diffuse and (c) global horizontal irradiance for the test dataset. Direct and diffuse irradiance are normalized by dividing by clear-sky direct irradiance, global horizontal irradiance is normalized by dividing by clear-sky irradiance. spp = samples per pixel (per spectral quadrature point).}
\label{pdf_plot}
\end{figure}

To quantitatively compare spatial irradiance distributions across all fields of the test dataset, we compare probability density functions (PDFs) of normalized direct, diffuse and global horizontal irradiance for the entire test dataset (Figure \ref{pdf_plot}). The PDF of direct irradiance has two distinct modes, the left peak corresponding to cloud shadows and the right peak corresponding to sunlit areas (Figure \ref{pdf_plot}a). With the denoising approach, the target cloud-shadow peak is approached near exactly for both 1 and 4 spp. The target peak of the sunlit mode is more closely matched by the denoising approach than the 128 spp reference (Figure \ref{pdf_plot}a). The peak is still slightly over-smoothed, caused mostly by the errors on shadow edges (Figure \ref{dir_fields}f), but much more accurate than the noisy input.

The PDF of diffuse irradiance also has two peaks (Figure \ref{pdf_plot}b), where the left peak corresponds to small cloud cover and the right peak to large cloud cover. The target diffuse PDF is reproduced almost exactly by the denoising approach (Figure \ref{pdf_plot}b), while the 128 spp reference underestimates both peaks slightly. The global horizontal irradiance and its corresponding PDF (Figure \ref{pdf_plot}c) is obtained by adding up separately denoised direct and diffuse irradiance fields. The target GHI distribution is also approximated very closely by the denoising approach, for both 1 and 4 spp.

\subsection{Denoising Efficiency Across Sampling Budgets}
\begin{figure}[h]
\noindent\includegraphics[width=\textwidth]{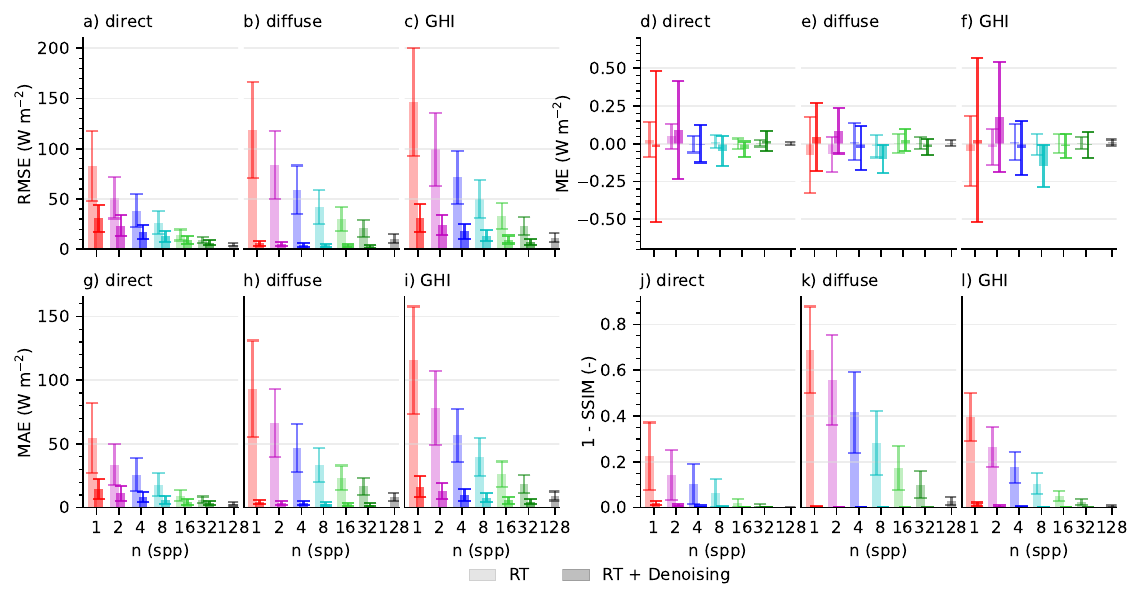}
\caption{(a,b,c) Root Mean Squared Error (RMSE), (d,e,f) Mean Error (ME), (g,h,i) Mean Absolute Error (MAE) and 1$-$SSIM (j,k,l) for direct, diffuse and global horizontal irradiance (GHI), comparing ray tracer input (RT) for different sampling budgets (n) with corresponding denoising performance (RT + Denoising). Error bars display standard deviations of the error metrics across the test dataset. spp = samples per pixel (per spectral quadrature point).}
\label{rmse_me_plot}
\end{figure}

For diffuse irradiance, the gain in accuracy due to denoising is largest with 1 sample per pixel (Figure \ref{rmse_me_plot}b). The RMSE and MAE of the input decrease with 95\%, while the mean bias and standard deviation of the mean error remain relatively small (Figure \ref{rmse_me_plot}e). The RMSE of the noisy diffuse input decreases from 119 to 21 $\mathrm{W\ m^{-2}}$ when increasing the sampling budget of the ray tracer from 1 to 32 spp (following RMSE $\propto \frac{1}{\sqrt{n}}$), while the RMSE after denoising decreases only from 5.8 to 2.9 $\mathrm{W\ m^{-2}}$. Since the errors are already relatively low, increasing the sampling budget for diffuse irradiance may not be worth the additional computational costs. Denoising diffuse irradiance from 1 spp outperforms the 128 spp reference regarding RMSE, MAE and 1$-$SSIM (Figure \ref{rmse_me_plot}e).

Also for direct irradiance, the relative benefit of the denoising approach is largest at 1 spp, decreasing the RMSE of the input with 63\%. However, the absolute magnitude of the RMSE and MAE are still considerably high (Figure \ref{rmse_me_plot}a). Using a larger sampling budget to denoise direct irradiance decreases absolute errors, but reduces the benefits of the denoiser (Figure \ref{rmse_me_plot}a). The RMSE and MAE of the 128 spp reference are lower than the errors of the denoising approach for all sampling budgets: in terms of accuracy, a larger sampling budget outperforms denoising for direct irradiance. Still, the low 1$-$SSIM scores (Figure \ref{rmse_me_plot}j) for the denoising approach indicate that the autoencoder is able to reproduce the most relevant structural information in the direct irradiance fields.

Summing the separately denoised direct and diffuse irradiance to GHI shows that denoising from 8 samples per pixel results in similar performance compared to the 128 spp reference. Although the average RMSE of the denoising approach is 1.9 $\mathrm{W\ m^{-2}}$ higher (Figure \ref{rmse_me_plot}c), both the MAE and 1$-$SSIM are lower. 

Denoising increases the standard deviation of the mean errors for direct irradiance compared to the ray tracer input (Figure \ref{rmse_me_plot}d), although the errors remain relatively small compared to the mean radiative fluxes. The mean errors of direct and diffuse do not necessarily cancel each other out, resulting in a somewhat larger bias in GHI (Figure \ref{rmse_me_plot}f).

The spread in RMSE is primarily explained by differences in zenith angle and cloud field characteristics. In general, the RMSE decreases with increasing zenith angle because total irradiance and hence absolute errors decrease. Smaller cloud cover results in a smaller diffuse fraction, decreasing the absolute errors of diffuse irradiance. Additionally, cloud fields with larger cloud structures have a smaller fraction of pixels on the edges of cloud shadows, resulting in lower field-averaged errors for direct irradiance.

\begin{figure}[h]
\noindent\includegraphics[width=\textwidth]{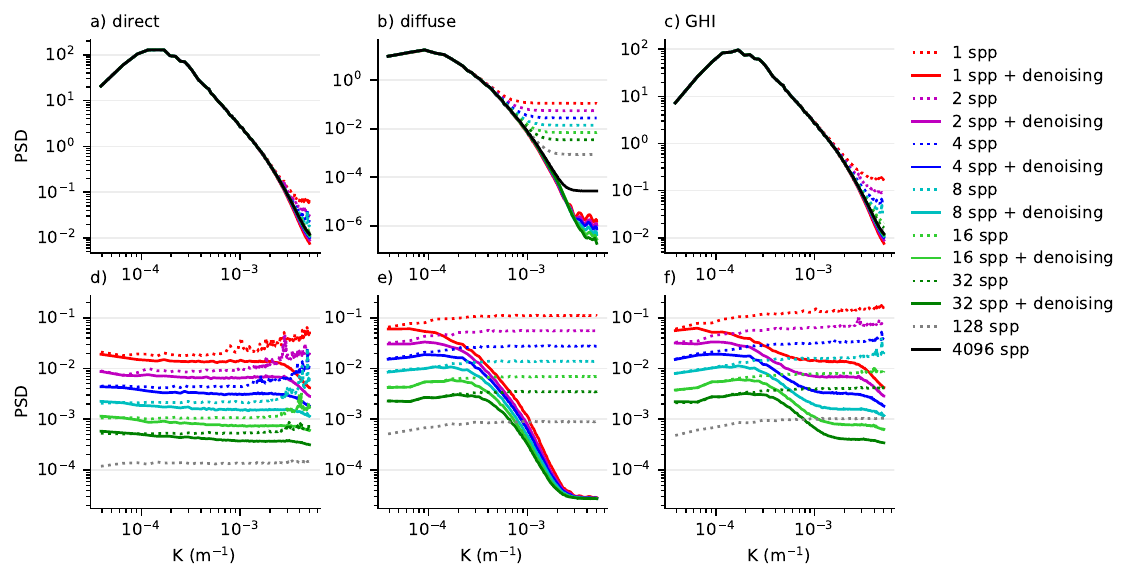}
\caption{Radially-averaged power spectral density (PSD) of the surface irradiance fields (top row) and errors with respect to the 4096 spp target (bottom row) for (a,d) direct, (b,e) diffuse and (c,f) global horizontal irradiance (GHI). Spectra are averaged over all irradiance fields in the test data.}
\label{spectra}
\end{figure}

Power spectra of the solar irradiance fields further illustrate the performance of the denoiser (Figure \ref{spectra}a-c). Most of the energy is contained at large spatial scales (L $\ge$ O(1 km)), irrespective of the sampling budget, but lowering the sampling budget increases the energy at the smallest scales. The denoising algorithm preserves the energy at large scales, while removing the energy from the smaller scales. The denoised irradiance fields even contain less energy at the smallest scales than the 4096 spp target, indicating that the denoiser does not overfit to remaining noise patterns in the target. However, there is also a risk of smoothing out small-scale irradiance fluctuations that are caused by small-scale atmospheric inhomogeneities. 

The power spectra of the irradiance errors with respect to the 4096 spp target (as in Figure \ref{dif_fields}f and \ref{dir_fields}f) increase approximately uniformly over all wave numbers when the sampling budget decreases (as in Figure \ref{spectra}d-f). Denoising reduces the error at small spatial scales, while the error at larger scales mostly remains.
The contribution of these errors at larger scales (L $\ge$ O(1 km)) to the total variance is negligible, however.
At small scales (L $\le$ O(1 km)), the spectral energy is reduced below the 128 spp reference for diffuse irradiance (Figure \ref{spectra}e), but not for direct irradiance (Figure \ref{spectra}d). This corroborates our earlier results, showing that denoising is more efficient for diffuse than for direct irradiance.   

In theory, the runtime increases linearly with the number of samples. However, the runtimes of the 128 spp (5.02 s) and 8 spp (0.79 s) and 4 spp (0.64 s) simulations are on average only about 11, 1.8, and 1.5 times slower than the 1 spp (0.43 s) simulation, because the GPU is not used optimally for low sampling budgets. For higher resolutions, larger domains, or lower-performance GPUs, the GPU will be more fully utilized for low sampling budgets, enlarging the differences in runtimes and increasing the computational benefits of the denoising approach. If the GPU is used to its full extent, the runtime does indeed increase linearly: 4096 spp takes 160.1 s, about 32 times more than 128 spp. 

\subsection{Importance of auxiliary variables}
\begin{figure}[h]
\noindent\includegraphics[width=\textwidth]{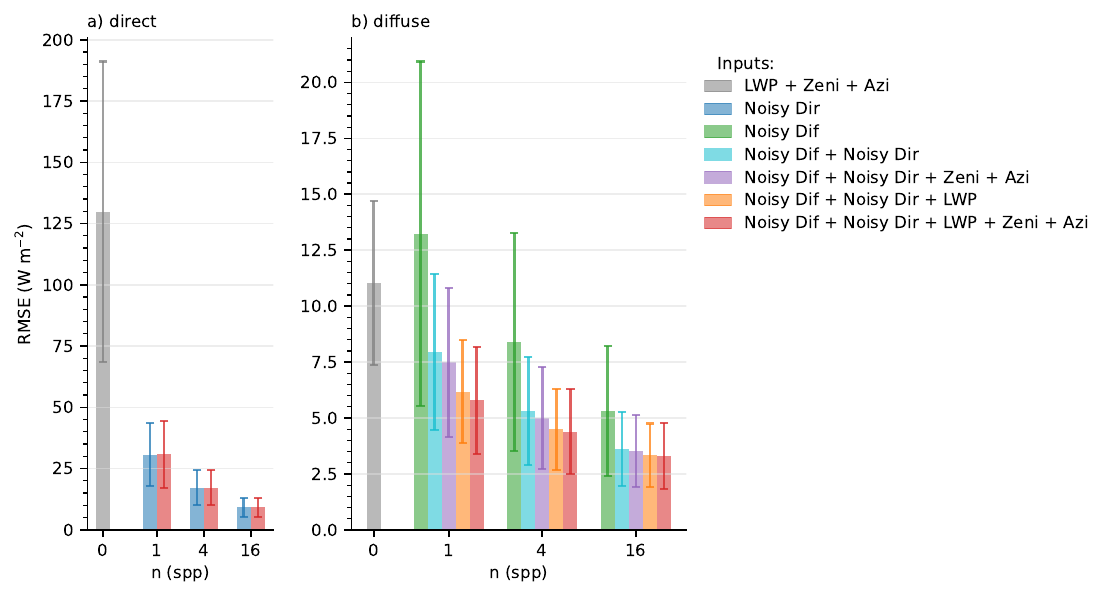}
\caption{Root Mean Squared Error (RMSE) for (a) direct and (b) diffuse irradiance for algorithms trained on a selection of the input features. Error bars display standard deviations of the RMSE across the test dataset. Noisy Dir/Dif = direct/diffuse irradiance with low sampling budgets (see x-axis); Zeni = solar zenith angle; Azi = solar azimuth angle; LWP = Liquid Water Path; spp = samples per pixel (per spectral quadrature point).}
\label{auxvar_plot}
\end{figure}

To understand the contribution of the auxiliary variables, we retrained various models with only a selection of the input features (Figure \ref{auxvar_plot}). Interestingly, the algorithms denoising direct irradiance do not improve when including auxiliary variables alongside the noisy direct irradiance (Figure \ref{auxvar_plot}a). Apparently, the noisy direct irradiance field contains sufficient information regarding locations of cloud shadows and magnitude of direct incoming radiation in sunlit areas, making the additional information on cloud locations and solar angles redundant. 

Using only the liquid water path (LWP) and solar angles as input and excluding noisy irradiance (hence 0 spp) deteriorates the accuracy of the denoiser, although performance might be improved by adding more cloud information such as top and base heights as direct radiation is mostly a geometric problem.
Predicting diffuse irradiance with only the LWP and solar angles is remarkably accurate (Figure \ref{auxvar_plot}b). Using this approach, there would be no need to generate noisy irradiance fields, which would greatly reduce computational costs. However, the bias of this algorithm is larger ($\mathrm{\overline{ME}}=0.36$, $\mathrm{\sigma_{ME}}=2.99$ W m$^{-2}$) than for algorithms including noisy irradiance (e.g. using only 1 spp diffuse irradiance as input: $\mathrm{\overline{ME}}=-0.09$, $\mathrm{\sigma_{ME}}=0.32$ W m$^{-2}$). Also, the algorithm using only LWP and solar angles as input is likely highly overfitted to irradiance patterns associated with shallow cumulus clouds.

The accuracy of diffuse irradiance significantly improves when auxiliary variables are included in the input along with the noisy diffuse irradiance. For 1 and 4 spp, the RMSE more than halves when including all additional variables (Figure \ref{auxvar_plot}b). The contribution of auxiliary variables decreases with the sampling budget, because the initial irradiance fields hold more information. The most important auxiliary variable for the diffuse models is the noisy direct irradiance, providing information on the location of cloud shadows on the surface. The liquid water path (LWP) decreases the RMSE with an additional 2 $\mathrm{W\ m^{-2}}$ for a sampling budget of 1 spp, while the solar angles contribute only marginally to an improvement in performance, most likely because this information is already implicitly contained in the noisy direct irradiance input.

\section{Conclusions and Discussion}
Monte Carlo ray tracing is an accurate method to compute radiative transfer, but computationally expensive because large sampling budgets are required for sufficient convergence. Here, we use a denoising autoencoder to efficiently infer well-converged surface solar irradiance fields, allowing to avoid having to run the ray tracer with large sampling budgets. We demonstrate that the denoising approach is effective when limiting diffuse irradiance sampling to 1 sample per pixel (spp), which is approximately twice as accurate and an order of magnitude faster than a reference using 128 spp. Denoising of direct irradiance yields good performance in sunlit and shaded areas, while persisting errors on the edges of cloud shadows cause larger field-averaged errors compared to the 128 spp reference. Furthermore, we show that auxiliary inputs halve errors for diffuse irradiance for a sampling budget of 4 spp or less, but do not improve prediction accuracy for direct irradiance.

The remaining noise on cloud shadow edges and inability to use information from additional variables may argue for another approach regarding computation of direct irradiance. 
Tilted independent column approximations, which apply 1D radiative transfer methods along slanted columns following the path of the incoming sun rays, may provide an efficient alternative for three-dimensional direct radiative transfer \cite<e.g.,>{Gabriel1996, Varnai1999, Wapler2008,Wissmeier2013}. 
As estimating 3D diffuse radiative effects from 1D radiative transfer computations comes with the difficulty of parameterizing appropriate convolution kernels \cite{Marshak1995,Zuidema1998,Wissmeier2013,tijhuis2023}, Monte Carlo ray tracing with denoising can then still be used to obtain accurate diffuse irradiances. 

While we find a small mean bias in the noisy irradiance fields of the test data, this bias converges to zero when averaging over more fields since Monte Carlo methods are by definition unbiased. This property should ideally be preserved after denoising, yet we find that denoising causes mean errors to drift farther from zero. Although remaining biases after denoising are still relatively small ($<$ 0.2 W m$^{-2}$), their implications need further attention. Possibly, the bias of the denoiser can be reduced by increasing the number of training samples or by enforcing conservation of mean irradiance more locally. Future work should test on which temporal and spatial scales biases remains, and if this remaining biases are significant enough to impose conservation of mean irradiance more strictly in future denoising algorithms.  

To enable wide-scale implementation of the denoiser, its applicability should be improved by training with a wider range of horizontal resolutions and a greater diversity of cloud types. Furthermore, additional features in the input may improve denoising accuracy. For instance, the layer depth and base height of clouds can vary for an equal liquid water path, with significant impact on the spatial distribution of irradiance \cite{He2024}. Other potentially relevant features that are linked to surface irradiance include the surface albedo \cite{Villefranque2023}, aerosol optical depth \cite{gristey2022}, cloud droplet number concentration \cite{Gristey2020b}, ice water path, and total column water vapor.

To bring our approach to 3D atmospheric models, we need to extend the denoising from two to three dimensions, and deal with the time dimension. Denoising 3D fields of radiative heating rates strongly increases the complexity of our denoiser, but at the same time the amount of denoising needed might be limited, as the sensitivity of atmospheric simulations to noise in radiative heating rates could be small \cite{Pincus2009}. The application of the denoiser in successive time steps provides an opportunity for recurrent connections \cite{Chaitanya2017} that link hidden layers from the previous time step to hidden layers of the current time step to propagate information on the structure of heating rates and surface irradiances in time.
Therefore, despite some challenges that still need to be overcome, our denoising approach  brightens the perspective of bringing 3D ray tracing to operational models.

\section*{Open Research Section}
All data and scripts, as well as the version of the radiative transfer code used for this paper, are available on Zenodo \cite{Reeze2024}

\acknowledgments
We acknowledge funding by the Nederlandse Organisatie voor Wetenschappelijk Onderzoek (grant nos. VI.Vidi.192.068, NWO-2023.003).
Simulations were carried out on the Snellius Dutch national supercomputer of SURF.

\bibliography{references}

\end{document}